% MNDSAMPL.TEX   last modification: 8.12.92
% A sample two column Monthly Notices article.
%
% Marginal adjustments using \pageoffset maybe required when printing
% proofs on a Laserprinter (this is usually not needed).
% Syntax: \pageoffset{ +/- hor. offset}{ +/- vert. offset}
% e.g.    \pageoffset{-3pc}{-4pc}
\input mnrass.sty
\pageoffset{-2.0pc}{-2.5pc}

%\def\etal{{\it et al.}}

% \font\euler=eurm10
% \def\umu{\hbox{$\euler\mu$}} \def\upi{\hbox{$\euler\pi$}}
% Uncomment the above two lines if the Euler font is available

%\Referee   %  uncomment this for referee mode
\Autonumber  %  auto-number sections, subsections and subsubsections

% \pagerange, \pubyear and \volume are defined at the Journals office and
% not by an author.

\pagerange{000--000}
\pubyear{1997}
%\volume{000}
%\umufiche{000}     % for articles with microfiche
%\authorcomment{Based on observations obtained at the European Southern
%Observatory, La Silla, Chile}  % author comment for footline

\begintopmatter  %  start the two spanning material

\title{Cepheids in NGC 1866 - a Test for Pulsational Models}

\author{Giuseppe Bono$^1$ \& Marcella Marconi$^2$}
\affiliation{$^1$Trieste Astronomical Observatory, Via G. B. Tiepolo 11,
34131 Trieste, Italy - E-Mail: bono@oat.ts.astro.it} 

\affiliation{$^2$Dept. of Physics, University of Pisa, Piazza Torricelli 2,
56100 Pisa, Italy - E-Mail: marcella@astr1pi.difi.unipi.it}

\shortauthor{G. Bono \& M. Marconi}
\shorttitle{Cepheids in NGC 1866 as a test for pulsational models}

% \acceptedline is to be defined at the Journals office and not
% by an author.

%\acceptedline{Accepted 1994 February 30. Received 1994 February 29;
%  in original form 1994 February 28}

\abstract
\tx
We present nonlinear theoretical results concerning the predicted pulsational
behaviour of stellar models suitable for the sample of 
Cepheids belonging to the Large Magellanic Cloud (LMC) star cluster 
NGC 1866. The blue and red edges of the instability strip, transformed 
into the observative plane (V, B-V) by adopting current values for the 
distance modulus (DM=18.57 mag., Welch et al. 1991) and for the reddening 
(E(B-V)=0.06 mag., Arp 1967; Walker 1974; Walker 1987), are in  
agreement with the observed distribution of Cepheids in the quoted cluster. 
Moreover, 
the distribution of observational data in the Bailey diagram (V amplitude 
versus period) appears in agreement with theoretical predictions derived 
adopting the luminosity level predicted by canonical evolutionary models. 
The role played by the direct {\em ab initio} integration of the coupling 
between pulsation and convection and by the input physics on the 
smoothness of light and velocity curves is briefly discussed.

\noindent 
\keywords Cepheids -- Magellanic Clouds -- Globular Clusters: 
individual (NGC 1866) -- stars: oscillations  
\maketitle  %  finish the two spanning material

\section{INTRODUCTION}
\tx 

\noindent 
The LMC globular cluster NGC 1866 has been already used extensively for 
testing stellar evolutionary models and for shedding light on some problems 
currently debated in the literature (Chiosi et al. 1989; 
Brocato, Castellani \& Piersimoni 1994). 
This cluster is also expected to play a key role in the theory of 
stellar pulsation, since  it provides a beautiful sample of 23 Cepheids
with a common value for both distance modulus and reddening.
Welch et al. (1991, hereinafter referred to as WCFMM) already disclosed
that this sample of radial pulsators  shows a well defined instability 
strip which is not revealed by current data on Galactic
Cepheids, and represents a fundamental benchmark for both modal stabilities 
(fundamental and first overtone) and pulsational amplitudes predicted by 
theoretical models. 

\noindent 
As a first step of an extended and detailed investigation of Cepheid 
hydrodynamical models we use NGC 1866 to test both the accuracy and the 
internal consistency of the theoretical framework we developed  
for evaluating  the approach to limit cycle stability and the predicted 
pulsational amplitudes.
The main aim of this paper is to discuss the comparison between a new 
set of limiting amplitude, nonlinear, nonlocal and time-dependent convective 
models and the observational constraints provided by this cluster. 
In \S 2 the theoretical results are briefly discussed while \S 3 presents  
the comparison with the sample of Cepheids in 
NGC 1866 for which magnitudes, colours, and pulsational amplitudes are 
currently available. 
The last section summarizes the main results of this investigation and 
outlines future theoretical and observational perspectives.

\section{THEORETICAL FRAMEWORK}

\tx
According to current evolutionary predictions for Cepheids in NGC 1866, 
we adopted a fixed stellar mass ($M/M_{\odot}$=5.0). 
Even though no direct estimation of the metallicity of this 
cluster has been provided yet, we adopted the metal abundance obtained  
by Russell \& Bessell (1989) and Russel \& Dopita (1990) for young LMC 
stars, i.e. Z=0.008. We also adopted the helium content Y=0.25 provided by 
Dufour (1984) on the basis of He abundance evaluations in H II regions.   
Similar physical parameters have also been assumed by Chiosi et al. 
(1992,1993 and references therein) in their extensive 
investigation of the pulsational properties of Cepheids linear models.  
For properly evaluating the topology of the instability strip we 
computed four sequences of both linear and nonlinear models located at 
different luminosity levels (log $L/L_{\odot}$ = 2.80, 3.07, 3.3, 3.6) 
which cover a wide effective temperature range ($4800 \le T_e \le 6700$ K). 
The location of the blue and red edges of fundamental and first overtone 
pulsators has been evaluated by adopting, close to these boundaries, a 
temperature step of 100 K. 
The theoretical framework adopted for evaluating the pulsational  
properties of Cepheids has already been described in a series of previous
papers (Bono \& Stellingwerf 1993,1994; Bono et al. 1997b and references 
therein). In this section we only discuss the main differences with 
the quoted investigations concerning the physical assumptions adopted to 
approach  Cepheids pulsational models. 

\bigskip

\noindent
For providing a good spatial resolution throughout the envelope 
model, the static structures have been computed by adopting an outer 
boundary optical depth $\tau= 0.0001$. At the same time, the mass 
ratio (h) between consecutive zones has been assumed equal to 1.04 in the 
stellar layers located at temperatures lower than $6.0\times10^4$ K, 
whereas for higher temperatures we adopted h=1.2.  
The inner boundary has been fixed so that the base of the envelope 
was approximately located at a distance from the stellar centre of
the order of 10\% of the photospheric radius (i.e. $r_0=0.1\times R_{ph}$). 
Unlike envelope models of low-mass stars, these assumptions
ensure that the mass included in a typical Cepheid model ranges from 40 to 
50 percent of the total stellar mass.  
Both linear and nonlinear models have been computed by adopting the 
radiative opacity tables recently provided by Iglesias \& Rogers (1996) 
for temperatures higher than 10,000 K,
and the molecular opacities provided by Alexander \& Ferguson (1994)
for lower temperatures. The 
method adopted for handling the opacity tables is described in Bono,
Incerpi \& Marconi (1996).  

\bigskip

\noindent 
The nonlinear analysis has been performed by perturbing the linear radial 
eigenfunction of the first two modes with a constant velocity amplitude 
of 10 km/sec. In contrast with nonlinear RR Lyrae and Type II 
Cepheid models, which have been computed  with  timesteps  
fixed by the CFL condition for properly handling the development and 
propagation of the shock front along the pulsation cycle, the nonlinear 
Cepheid models have been computed by adopting a number of timesteps per  
period which range from 300 for first overtone models to 400 for fundamental 
pulsators. This choice is due to the evidence that along a full cycle Cepheid
models develop only mild shocks in the hydrogen and helium ionization regions.
As a test, few selected models computed by adopting the CFL conditions do not 
present any substantial difference in the pulsation amplitudes in comparison 
with the models computed by adopting a fixed timestep.  
The number of pulsation cycles needed to approach the asymptotic amplitudes 
depends on the location of the model inside the instability strip and 
ranges from 1,000 to more than 10,000.     

\bigskip
  
\noindent 
A thorough comparison of our models with both linear (Chiosi et al. 1993; 
Morgan \& Welch 1996) and nonlinear radiative Cepheid models 
(Sebo \& Wood 1995; Simon \& Kanbur 1995; Buchler et al. 1996), together 
with a detailed analysis of the physical structure of the envelope models 
will be discussed in a forthcoming paper (Bono \& Marconi 1997).    

\noindent 
The top panel of Figure 1 shows a selected sample of fundamental and first 
overtone bolometric light curves located at log $L/L_{\odot}$ = 3.07, 
i.e. at the luminosity level predicted by canonical evolutionary models. 
The first interesting result is that moving from the blue to the red edge 
of the instability 
strip the light curves show smooth changes along the pulsation cycle and 
do not present, at least in this period range, any peculiar bump and/or 
dip. The comparison with results available in the literature (Karp 1975; 
Carson \& Stothers 1984) strengthens the key role played by the inclusion 
of a time-dependent treatment of convective transport which reduces the 
pulsational amplitudes and at the same time ensures a smooth excursion 
of physical variables across the ionization regions.  
The bottom panel of Figure 1 shows the surface radial velocities of 
the same models shown in the top panel. 
The smoothness of these curves over the pulsation cycle 
is the result of a long {\em ab initio} integration 
time during which the radial motions approach their asymptotic 
amplitudes and the spurious high-order modes introduced by the initial 
perturbation settle down.  
\figure{1}{S}{80mm}{\bf Figure 1. {\it Top}: Selected bolometric light 
curves as a function of the pulsational phase. Solid and dashed lines 
are referred to first overtone and fundamental pulsators respectively. 
The magnitude scale is referred to the top curve; the other ones have been 
artificially shifted by -0.60 mag. The nonlinear periods are also labelled. 
{\it Bottom}: Surface radial velocity curves of the models plotted 
in the top panel. The velocity scale is referred to the top curve;  
the other ones have been artificially shifted by -35 km/sec. 
The effective temperatures are also labelled.} 
\noindent 
A more detailed summary of the nonlinear observables obtained in the 
present investigation for both fundamental and first overtone 
pulsators is given in Table 1. This table presents from left to right 
the logarithmic luminosity, the effective temperature, the nonlinear
pulsation period, and the fractional radius variation, i.e. 
$\Delta R/R_{ph}=(R^{max}\,-\,R^{min})/R_{ph}$, where $R_{ph} $ is the
photospheric radius. Column (5) reports the "pure" radial velocity 
amplitude of the surface zone, i.e. no correction factor for limb darkening 
has been taken into account. Column (6) gives the bolometric amplitude, 
while the two subsequent columns report the amplitude of static (7) and 
effective (8) logarithmic surface gravity. The static gravity is referred 
to the surface zone, whereas the effective gravity has been evaluated 
by adopting the method suggested by Bono, Caputo \& Stellingwerf (1994).  
Columns (9) and (10) give the amplitude of the surface temperature. 
The former is referred to the temperature of the outer boundary, 
whereas the latter is derived taking into account the surface variations 
over a full cycle of both luminosity and radius. The final column shows 
the pulsational amplitude in the V band (see \S 3 concerning the 
transformation into the observative plane).  

\noindent 
In order to point out {\em who is doing what} concerning the 
driving and the damping of the pulsation instability, Figure 2 shows the 
total work curves referred to a fundamental model centrally located 
in the instability strip (log $L/L_{\odot}$ = 3.07, $T_e$=5800 K). 
The light and velocity curves of this model are shown in Figure 1. 
Since we are interested in disclosing 
the destabilization effects of different elements (H, He, metals), the 
total work terms have been evaluated according to the  relation
$dW/d (log <T >)$, where  $<T >$ is the time average temperature  
per zone over a full cycle (see Bono et al. 1997a). 
Note that in the literature the total works are generally evaluated as a
function of the exterior mass (i.e. $dW/d (log M_e)$). 
According to Figure 2 we find that H and He, as expected, 
are the main destabilization sources whereas the Z-bump, at least for the 
assumed metallicity, barely affects the pulsation stability. On the other 
hand, the region located between the HeII ionization zone and the Z-bump 
is the main radiative damping region of the envelope.  
Moreover, the artificial viscosity pressure does not affect at all the 
physical structure of this model, since it is vanishing throughout the 
envelope. Turbulent and eddy viscosity 
pressure terms supply a non-negligible amount of dissipation in the 
hydrogen and helium ionization regions which in turn reduces the 
pulsational amplitudes. 

In order to easily identify the opacity bumps which cause the appearance 
of both driving and damping regions, the bottom panel of Figure 2 shows 
the opacity excursions throughout the envelope structure.  
\figure{2}{S}{80mm}{\bf Figure 2. {\it Top}: Full amplitude nonlinear 
total work curves versus the logarithmic time average temperature, surface 
at right. This fundamental model is centrally located in the instability 
strip ($log L/L_{\odot}$=3.07; $T_e$=5800 K). The solid line shows the 
total work, the dashed line the turbulent work (i.e. the work due to 
both turbulent and eddy viscosity pressure), and the dashed-dotted 
line the artificial viscosity work. Positive and negative areas denote 
driving and damping regions respectively.  
{\it Bottom}: Logarithmic opacity of the envelope as a function of the 
logarithmic time average temperature. The opacity profile has been plotted 
at each time step over a full cycle. The arrows mark the location of the 
main opacity sources.}  
The opacity bump which appears in the outermost regions is worth 
mentioning. This opacity increment, at densities typical of Cepheid 
surface layers, is mainly due to $H^-$ and partially to the Rayleigh 
scattering of hydrogen (see Fig. 7 in Alexander \& Ferguson 1994).   
In fact, during a large part of the pulsation cycle the surface layers 
of this model oscillate at temperatures lower than 5,000 K, reaching the 
maximum temperature excursion (T=5,250 K) only around the phases of 
the luminosity maximum. 

\noindent 
This feature, which has a negligible effect on the pulsation driving, 
could play a key role in the smoothness of both light and velocity 
curves. In fact, thanks to the inclusion of molecular opacities, the 
radial displacement of surface layers located above the hydrogen ionization 
region is governed by the physical excursion of temperature and density.

\section{COMPARISON WITH OBSERVATIONS}
\tx

\noindent 
In this section  theoretical results are compared with 
the photometric data for NGC 1866 provided by WCFMM. These data refer
to only nine out of the 23 Cepheids identified in NGC 1866, since 
photometry is not provided in the quoted paper for the remainder.
In a subsequent paper Welch \& Stetson (1993) applied their variable star
detection technique to several frames centered on NGC 1866 and provided
periods for all but one variable of the whole sample. 
However, both mean magnitudes and colours presented in that paper are 
affected by a poor photometric accuracy due to the crowding of central 
regions and thus are of no use.    

\noindent
Theoretical boundaries for both fundamental and first overtone 
instability strips 
 (Z=0.008, Y=0.25 and $M=5M_{\odot}$) have been transformed into the 
observative plane V-(B-V) by using Kurucz's bolometric corrections and 
colour-temperature relations (Kurucz 1992). In Figure 3 the transformed 
boundaries are shown together with the observed Cepheids of NGC 1866 
studied by WCFMM. Following WCFMM discussion we have 
assumed for the cluster a reddening value E(B-V)=0.06 mag. and a 
distance modulus DM=18.57 mag.
\figure{3}{S}{80mm}{\bf Figure 3. Fundamental and first overtone 
instability boundaries for Z=0.008, Y=0.25 and $M=5M_{\odot}$.  
The observational data are referred to Cepheids in NGC 1866. 
See text for further details.}
\noindent 
Solid lines represent, left to right, the First Overtone Blue Edge (FOBE), 
the Fundamental Blue Edge (FBE), the First Overtone Red Edge (FORE) and 
the Fundamental Red Edge (FRE). The FOBE and the FBE intersect at 
 $\log{L/{L_{\odot}}}\approx 3.4$ and therefore we expect that above 
 this luminosity level the instability strip is populated only by 
fundamental pulsators. Indeed at $\log{L/{L_{\odot}}}=3.6$ the nonlinear 
models computed by adopting the linear first overtone radial eigenfunction
do not show a stable limit cycle. On the other hand, toward lower  
luminosities, the FBE 
and the FORE intersect at $\log{L/{L_{\odot}}}=2.8$, thus removing 
completely the "OR region", i.e. the region of the instability strip where 
pulsators are characterized by a stable limit cycle in the fundamental and 
in the first overtone mode.

\bigskip

\noindent
The agreement with observational data appears satisfactory, since most 
of the observed variables are located well inside the theoretical 
instability strip around a magnitude level corresponding roughly to 
$\log{L/{L_{\odot}}}=3.07$. In agreement with early 
results obtained by Bertelli et al. (1993), the comparison also suggests 
the occurrence of first overtone pulsators among Cepheids in NGC 1866.  
The only deviant variable is HV 12204, already pointed out by WCFMM  
for its peculiar brightness and blueness. However, from the analysis of 
radial velocities these authors concluded that HV 12204 is not a cluster 
member. Two further features of the instability strip are worth mentioning:

1) moving from higher to lower luminosities the instability strip becomes 
narrower; in the investigated luminosity range 
the colour width roughly decreases of a factor of two as the outer edges  
of the instability region present different slopes. 
These findings support the results obtained by Fernie (1990) for Galactic 
Cepheids and the preliminary trend suggested by Olszewski (1995) for 
Cepheids in LMC clusters. 

2) The region of the instability strip in which only first overtones show a 
stable nonlinear limit cycle becomes larger at lower luminosities. 
This outcome supplies a straightforward but qualitative explanation 
for the large amount  
of first overtone pulsators -approximately 30 \% of the total sample- 
recently detected by microlensing experiments (Beaulieu et al. 1995; 
Cook et al. 1995) and supports the hypothesis originally suggested by 
B\"ohm-Vitense (1988) concerning the occurrence of such variables among 
short-period Cepheids.  

\bigskip

\noindent
In Figure 4 the theoretical relations between periods and V amplitudes
for both fundamental and first overtone models and for the labelled 
assumptions on luminosity are compared with the observational data
provided by WCFMM. Bolometric
amplitudes have been transformed into the V Johnson band by adopting
Kurucz's static atmosphere models. The V magnitude curve can be used to
derive either a magnitude-weighted amplitude or an intensity-weighted one,
which is subsequently converted into magnitude units. 
The differences between amplitudes obtained from these  two different
approaches turned out to be  negligible (less than 0.001 mag.). 
\figure{4}{S}{80mm}{\bf Figure 4. Comparison in the Bailey diagram between 
predicted and observed V amplitudes and periods. Symbols are referred to 
different luminosity levels (see labelled values). Fundamental and first 
overtone amplitudes are plotted by adopting solid and dashed lines 
respectively. Asterisks show the observed V amplitudes provided by WCFMM.} 

\noindent
It is worth noting that while the comparison in the HR diagram 
(instability strips) could be affected by uncertainties in the bolometric 
corrections, in the colour-temperature relations, in the reddening, and 
in the distance modulus, as well as in the method adopted to 
average the observed colours and magnitudes over the pulsation cycle, 
the comparison in the $A_V$-$\log{P}$ plane is independent of almost 
all these troublesome effects, and is therefore far more robust. 
As a consequence, we can infer that Cepheids in NGC 1866, at least those 
for which photometric data are given in the literature (WCFMM), 
 are well reproduced in the Bailey diagram by a sequence of pulsational 
models with Z=0.008, Y=0.25, $M=5M_{\odot}$ and 
$\log{L/L_{\odot}}\simeq{3.07}$, 
in remarkable agreement with the adopted evolutionary predictions. 
Note that the sequences of models characterized by the same values of 
stellar mass and chemical composition but by luminosity levels different 
from $\log{L/L_{\odot}}=3.07$ are clearly unable to match observational
data. This means that, in principle, the comparison in the period-amplitude
diagram proves to be a powerful instrument for estimating the luminosity 
level of cluster Cepheids.\par

A further interesting aspect worth being investigated in order to provide
both a suitable test of the present theoretical scenario and a tight 
constraint on the pulsation characteristics of classical Cepheids is the 
comparison between the Fourier parameters of theoretical and observed 
light curves (Simon \& Kanbur 1995). Even though such analysis is clearly
beyond the scope of this paper, a comprehensive comparison with 
observational data and the impact of the pulsation/convection interaction 
on observables will be discussed in a forthcoming paper (Bono \& 
Marconi 1997).

\section{CONCLUSIONS} 
\tx

\noindent 
The agreement between Cepheid theoretical models and the photometric 
data available in the literature provides a sound support to the 
plausibility of the physical assumptions adopted in the development 
of this new pulsational scenario. 
In a homogeneous theoretical context we derived both the blue and the 
red edges of the instability strip. In particular, it is worth noting 
that the red edges have been evaluated without invoking  any {\em ad hoc} 
assumption on the efficiency of the convective transport 
over the pulsation cycle. At the same time present limiting amplitude 
calculations strongly support the existence of first overtone pulsators 
among Cepheids. In fact, we found that there is a well defined region of 
the instability strip in which only first overtone pulsators present
a nonlinear limit cycle stability. 
On the other hand, taking into account the distribution of pulsational 
amplitudes in the Bailey diagram we find that the luminosity level 
of Cepheids along the blue loop is in agreement with the value 
predicted by canonical evolutionary models. 

\noindent
However, a deeper insight on the matter is prevented by the limited 
sample of Cepheids currently available. 
What is relatively surprising about photometric data of young LMC 
clusters (see Table 2 in Welch, Mateo \& Olszewski 1993 and references 
therein) with a good Cepheid sample is that even though they are the 
keystone for determining the 
P-L and the P-L-C relations on which rely the distance evaluations of the 
Local Group Galaxies, and in turn the calibration of secondary distance 
indicators,  we still lack a comprehensive observational scenario for 
these objects. 

\noindent 
Moreover, although several approaches have been recently undertaken 
for evaluating the metal abundance of stellar clusters in Magellanic Clouds 
(Geisler \& Mateo 1989; Olszewski et al. 1991), a detailed 
analysis of the chemical composition spread of Cepheids 
in Galactic and Magellanic Clouds clusters is even more urgent.   

\noindent 
As far as future theoretical developments are concerned, we plan 
to extend the present nonlinear pulsational scenario by taking into 
account different stellar masses and chemical compositions for properly 
disclosing the dependence of Cepheids behaviour on these astrophysical 
parameters.

\section*{Acknowledgments}

\tx 
It is a pleasure to thank V. Castellani and F. Caputo for many warm and 
enlightening discussions on Cepheids and intermediate-mass stars.

\section*{REFERENCES}
\bibitem Alexander, D. R., \& Ferguson, J. W. 1994, ApJ, 437, 879
\bibitem Arp, H. C. 1967, ApJ, 149, 91
\bibitem Beaulieu, J. P., et al. 1995, A\&A, 303, 137  
\bibitem Bertelli, G., Bressan, A., Chiosi, C., Mateo, M., \& Wood, P. R. 
1993, ApJ, 412, 160 
\bibitem B\"ohm-Vitense, E. 1988, ApJ, 324, L27  
\bibitem Bono, G., Caputo, F., Cassisi, S., Castellani, V., \& Marconi, M. 
1997a, ApJ, submitted  
\bibitem Bono, G., Caputo, F., Castellani, V., \& Marconi, M. 1997b, A\&AS, 
121, 327  
\bibitem Bono, G., Caputo, F., \& Stellingwerf, R. F. 1994, ApJ, L51  
\bibitem Bono, G., Incerpi, R., \& Marconi, M. 1996, ApJ, 467, L97
\bibitem Bono, G., \& Marconi, M. 1997, in preparation
\bibitem Bono, G., \& Stellingwerf, R. F. 1993, Mem. Soc, Astron. It., 64, 559
\bibitem Bono, G., \& Stellingwerf, R. F. 1994, ApJS, 93, 233
\bibitem Brocato, E., Castellani, V., \& Piersimoni. A. M. 1994, A\&A, 290, 59
\bibitem Buchler, J. R., Kollath, Z., Beaulieu, J. P., \& Goupil, M. J. 1996, 
ApJ, 462, 83L
\bibitem Carson, T. R., \& Stothers, R. B. 1984, ApJ, 276, 593
\bibitem Chiosi, C., Bertelli, G., Meylan, G., \& Ortolani, S. 1989, A\&A, 219, 167
\bibitem Chiosi, C., Wood, P., Bertelli, G., \& Bressan, A. 1992, ApJ, 387, 320
\bibitem Chiosi, C., Wood, P. R., \& Capitanio, N. 1993, ApJS, 86, 541
\bibitem Cook, K. H., et al. 1995, in IAU Colloq. 155, Astrophysical 
Applications of Stellar Pulsation, eds. R.S. Stobie \& P.A. Whitelock 
(San Francisco: ASP), 83, 221  
\bibitem Dufour, R. J. 1984, in IAU Symp. 108, Structure and Evolution of the 
Magellanic Clouds, ed. S. van den Bergh \& K.S. de Boer (Dordrecht: Reidel), 353
\bibitem Fernie, J. D. 1990, ApJ, 354, 295  
\bibitem Geisler, D., \& Mateo, M. 1989, RMxAA, 19, 101   
\bibitem Iglesias, C. A., \& Rogers, F. J. 1996, ApJ, 464, 943
\bibitem Karp, A. H. 1975, ApJ, 201, 641  
\bibitem Kurucz, R. L. 1992, in IAU Symp. 149, The Stellar Populations of 
Galaxies, eds. B. Barbury \& A. Renzini (Dordrecht: Kluwer), 225
\bibitem Morgan, S. M., \& Welch, D. L. 1996, preprint, astro-ph/9607068  
\bibitem Olszewski, E. W. 1995, in IAU Symp. 164, Stellar Populations, 
eds. P.C. van der Kruit \& G. Gilmore (Dordrecht: Kluwer), 181
\bibitem Olszewski, E. W., Schommer, R. A., Suntzeff, N. B., \& 
Harris, H. C.  1991, AJ, 101, 515  
\bibitem Russell, S. C., \& Bessel, M. S. 1989, ApJS, 70, 865
\bibitem Russell, S. C., \& Dopita, M. A. 1990, ApJS, 74, 93
\bibitem Sebo, K. M., \& Wood, P. R. 1995, ApJ, 449, 164
\bibitem Simon, N. R., \& Kanbur, S. M. 1995, ApJ, 451, 703
\bibitem Walker, M. F. 1974, MNRAS, 169, 199
\bibitem Walker, A. R. 1987, MNRAS, 225, 627
\bibitem Welch, D. L., Cote, P., Fisher, P., Mateo, M., \& Madore, B. F. 1991, 
AJ, 101, 490 (WCFMM) 
\bibitem Welch, D. L., Mateo, M., \& Olszewski, E. W. 1993, in IAU Colloq. 139,
New Perspectives on Stellar Pulsation and Pulsating Variable Stars, eds.
J.M. Nemec \& J.M. Matthews (Cambridge: Cambridge Univ. Press), 359  
\bibitem Welch, D. L., \& Stetson, P. B. 1993, AJ, 105, 1813
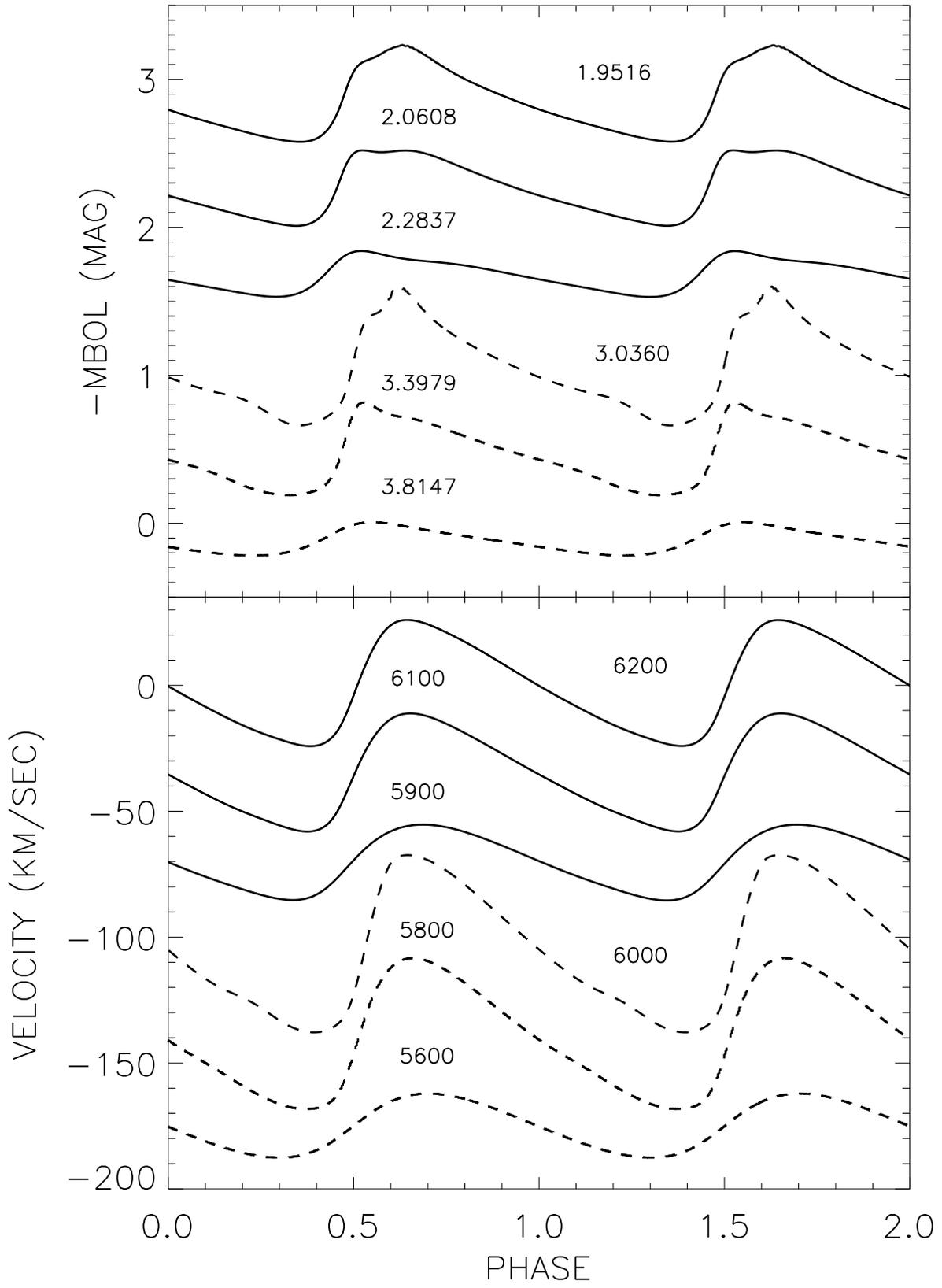
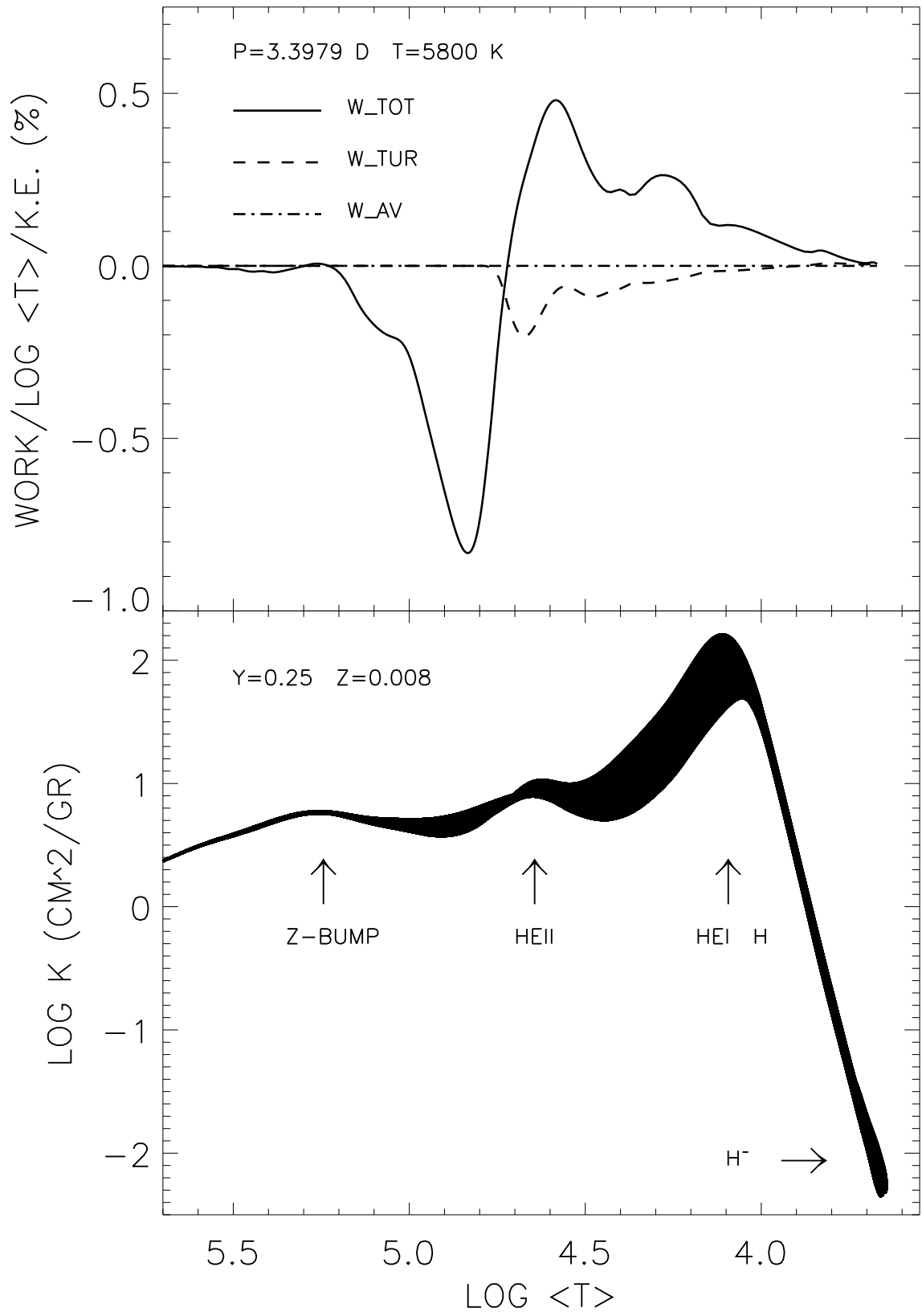
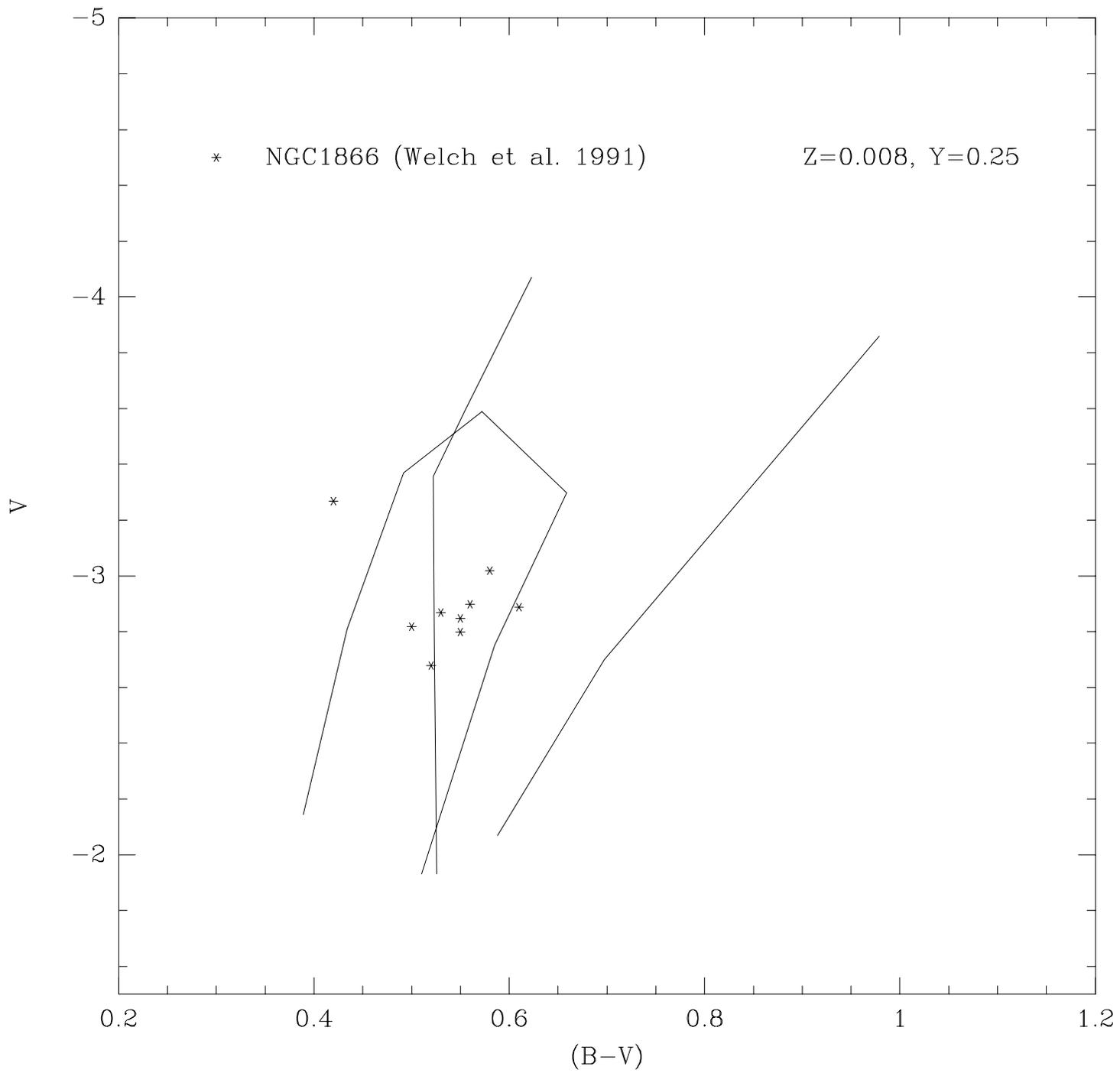
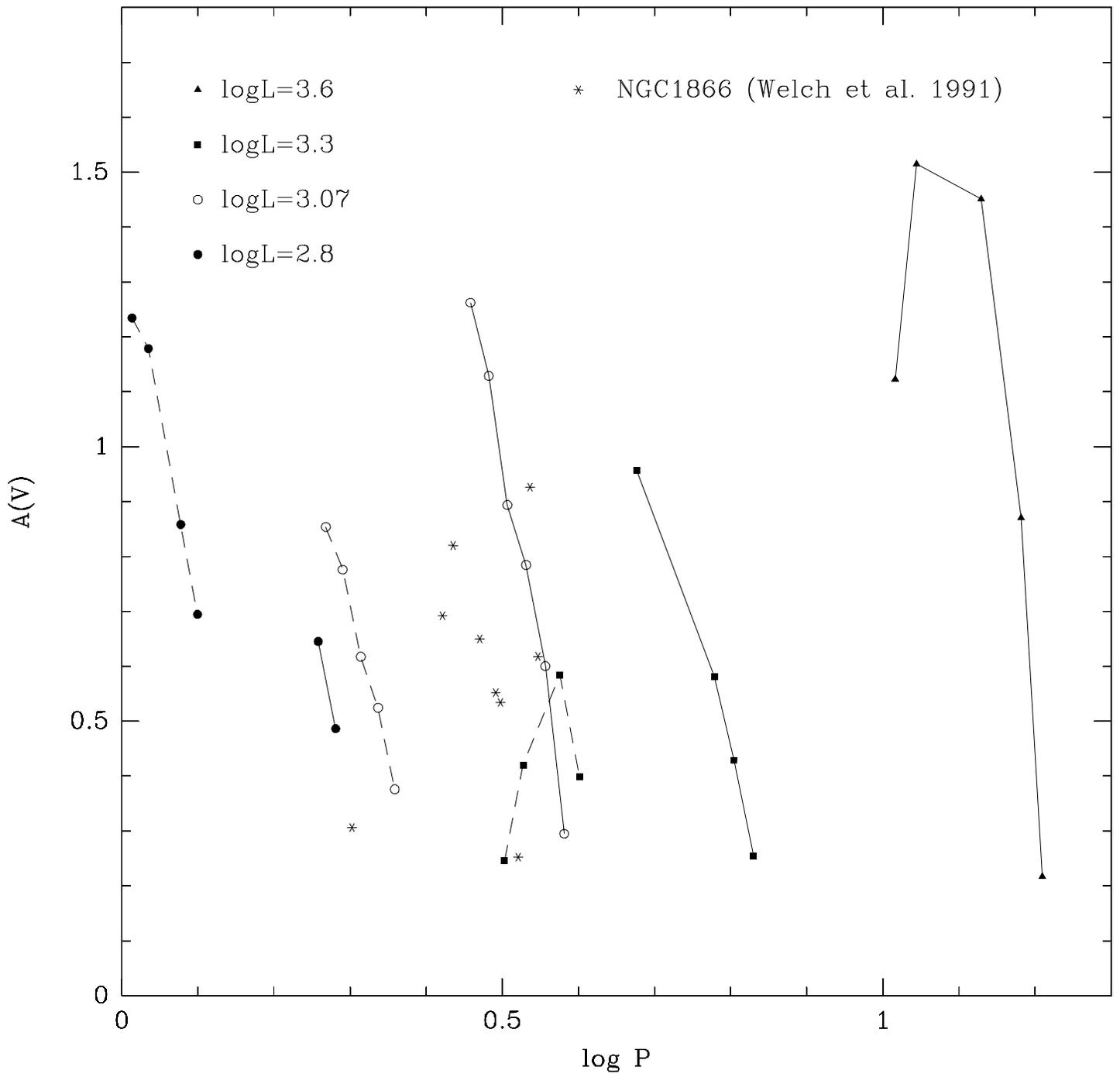
\bye